\documentclass[openacc]{rstransa}

\usepackage[draft]{changes}
\usepackage{mathtools}
\usepackage{dsfont}
\usepackage{soul,color}
\usepackage{comment}

\hypersetup{draft}

\newcommand{\ie}{\textit{i.e.}}
\newcommand{\eg}{\textit{e.g.}}

\begin{document}

\title{A statistical model of COVID-19 testing in populations: effects
  of sampling bias and testing errors}

\author{
Lucas B\"ottcher$^{1,2}$, Maria R. D'Orsogna$^{3,1}$ and Tom Chou$^{1,4}$}

\address{$^{1}$Dept.~of Computational Medicine, University of
  California, Los Angeles, 90095-1766, Los Angeles, CA, United
  States\\ 

$^{2}$Computational Social Science, Frankfurt School of Finance and
  Management, 60322 Frankfurt am Main, Germany

$^{3}$Dept.~of Mathematics, California State University at Northridge,
  Los Angeles, 91330-8313, CA, United States\\
  
$^{4}$Dept.~of Mathematics, University
  of California, Los Angeles, 90095-1766, Los Angeles, CA, United
  States}

\subject{Epidemiology, Medical Statistics, Applied Mathematics}

\keywords{COVID-19, testing, combinatorics}

\corres{Lucas B\"ottcher\\
\email{lucasb@g.ucla.edu}}

\begin{abstract}
We develop a statistical model for the testing of disease prevalence
in a population. The model assumes a binary test result, positive or
negative, but allows for biases in sample selection and both type I
(false positive) and type II (false negative) testing errors. Our
model also incorporates multiple test types and is able to distinguish
between retesting and exclusion after testing. Our quantitative
framework allows us to directly interpret testing results as a
function of errors and biases. By applying our testing model to
COVID-19 testing data and actual case data from specific
jurisdictions, we are able to estimate and provide uncertainty
quantification of indices that are crucial in a pandemic, such as
disease prevalence and fatality ratios.
\end{abstract}

\maketitle

\section{Introduction}
Real-time estimation of the level of infection in a population is
important for assessing the severity of an epidemic as well as for
guiding mitigation strategies. However, inferring disease prevalence
via patient testing is challenging due to testing inaccuracies,
testing biases, and heterogeneous and dynamically evolving populations
and severity of the disease.

There are two major classes of tests that are used to detect previous
and current SARS-CoV-2
infections~\cite{CDC_tests_overview}. Serological, or antibody, tests
measure the concentration of antibodies in infected and recovered
individuals. Since antibodies are generated as a part of the adaptive
immune system response, it takes time for detectable antibody
concentrations to develop. Serological tests should thus not be used
as the only method to detect acute SARS-CoV-2 infections. An
alternative testing method is provided by viral-load or antigen tests,
such as reverse transcription polymerase chain reaction (RT-PCR),
enzyme-linked immunosorbent assay (ELISA), and rapid antigen tests,
which are able to identify ongoing SARS-CoV-2 infections by directly
detecting SARS-CoV-2 nucleic acid or antigen.

Test results are mainly reported as binary values (0 or 1, negative or
positive) and often do not include further information such as the
cycle threshold ($\mathrm{Ct}$) for RT-PCR tests. The cycle threshold
$\mathrm{Ct}$ defines the minimum number of PCR cycles at which
amplified viral RNA becomes detectable. Large values of $\mathrm{Ct}$
indicate low viral loads in the specimen. An increase in $\mathrm{Ct}$
by a factor of about $3.3$ corresponds to a viral load that is about
one order of magnitude lower~\cite{tom2020interpret}. Cycle threshold
cutoffs are not standardized across jurisdictions and range from
values between $37-40$, making it difficult to compare RT-PCR test
results~\cite{mandavilli2020your}. Lower $\mathrm{Ct}$ cutoffs in the
range of $30-35$ may be more reasonable to avoid classifying
individuals with insignificant viral loads as
positive~\cite{mandavilli2020your}. 

Further uncertainty in COVID-19 test results arises from different
type I errors (false positives) and type II errors (false negatives)
that are associated with different assays. Note that inherent
  to any test, the threshold (such as $\mathrm{Ct}$ mentioned above)
  may be tunable. Therefore, besides intrinsic physical limitations,
  binary classification of ``continuous-valued'' readouts
  (\textit{e.g.}, viral load) may also lead to an overall error of
  either type \cite{PATRONE}.  In this work, we will assume that there
  is a standardized threshold and the test readout is binary; if any
  virus is detected, the test subject is positive. We will not
  explicitly model the underlying statistics of the errors but assume
  that the test readouts are binary but can be erroneous at specified
  rates. Some uninfected individuals will be wrongly classified as
infected with rate $\mathrm{FPR}$ and some infected individuals will
be wrongly classified as uninfected with rate $\mathrm{FNR}$. For
serological COVID-19 tests, the estimated proportions of false
positives and false negatives are relatively low, with
$\mathrm{FPR}\approx 0.02-0.07$ and $\mathrm{FNR}\approx
0.02-0.16$~\cite{fda_serological,cohen2020diagnosing,watson2020interpreting,bastos2020diagnostic}.
The $\mathrm{FNR}$s of RT-PCR tests depend strongly on the actual
assay method~\cite{lassauniere2020evaluation,whitman2020test} and may
be significantly larger than those of serological tests. Typical
values of $\mathrm{FNR}$ for RT-PCR tests lie between 0.1 and
0.3~\cite{fang2020sensitivity,wang2020detection} but might be as high
as $\mathrm{FNR}\approx 0.68$ if throat swabs are
used~\cite{wang2020detection,watson2020interpreting}. False-negative
rates may also vary significantly depending on the time delay between
initial infection and testing~\cite{bastos2020diagnostic}. According
to a systematic review~\cite{arevalo2020false} that was conducted
worldwide, the initial value of $\mathrm{FNR}$ is about 0.54,
underlying the importance of retesting.  Similar to serological tests,
reported false-positive rates of RT-PCR tests are about
$\mathrm{FPR}=0.05$~\cite{watson2020interpreting}.

Estimates of disease prevalence and other surveillance
metrics~\cite{bottcher2101using,bottcher2020case} need to account for
$\mathrm{FPR}$s and $\mathrm{FNR}$s, in particular if reported
positive-testing rates~\cite{CDC} are in the few percent range and
potentially dominated by type I errors. In addition to type-I/II
testing errors, another confounding effect is biased
testing~\cite{fenton2020covid}, that is, preferential testing of
individuals that are expected to carry a high viral load (\eg,
symptomatic and hospitalized individuals). Biasing testing towards
certain demographic and risk groups leads to additional errors in
disease prevalence estimates that need to be corrected for.

To account for type-I/II errors, bias, retesting and exclusion after
testing, we develop a corresponding framework for disease testing in
Sec.~\ref{sec:testing_model}. We apply our testing model to COVID-19
testing and case data in Sec.~\ref{sec:covid_data} and estimate
testing bias by comparing random-sampling testing
data~\cite{streeck2020infection} with officially reported, biased
COVID-19 case data in Sec.~\ref{sec:bias}. We conclude our study in
Sec.~\ref{sec:conclusion}.
\section{Statistical testing model}
\label{sec:testing_model}
Here, and in the following subsections, we develop a general
  statistical model for estimating the number of infected individuals
  in a jurisdiction by testing a sample population. The relevant
  variables and parameters to be used in our derivations are listed
  and defined in Table~\ref{tab:testing_variables}.
\begin{table}[htb]
\renewcommand*{\arraystretch}{1.6}
\begin{tabular}{| >{\centering\arraybackslash} m{10em}| 
>{\centering\arraybackslash} m{30em}|}\hline
symbol & definition
\\[1pt] \hline\hline
\,\,\, $N\in \mathds{Z}^{+}$\,\, & population in jurisdiction\\[1pt]  \hline
 \,\,\, $Q: [0,N]$\,\, & number of tests administered \\[1pt]  \hline
 \,\,\, $Q^+: [0,Q]$\,\, & recorded positives under error-free testing  \\[1pt]  \hline
 \,\,\, $\tilde{Q}^+: [0,Q]$\,\, & recorded positives under error-prone testing  \\[1pt]  \hline
\,\,\, $\displaystyle{f: [0,1]}$\,\, & 
\shortstack{\vspace{1em}\\ true proportion of infected individuals} \\[1pt]  \hline
\,\,\, $f_{\rm b}\!\coloneqq Q^{+}/Q: [0,1]$\,\, & 
\shortstack{\vspace{1em}\\ fraction of positives under biased, error-free testing}
\\ \hline
 \,\,\, $\displaystyle \tilde{f}_{\rm b}\!\coloneqq 
\tilde{Q}^{+}/Q: [0,1]$\,\, & 
\shortstack{\vspace{1em}\\ fraction of positives under biased, error-prone testing} 
\\ \hline
 \,\,\, $b\in \mathds{R}$\,\, & testing bias parameter \\[1pt]  \hline
 \,\,\, $\hat{b}, \hat{f}: [0,1]$\,\, & estimates of bias and underlying infection fraction \\[1pt]  \hline
 \,\,\, ${\rm FPR}: [0,1]$\,\, & false positive rate \\[1pt]  \hline
 \,\,\, ${\rm FNR}: [0,1]$\,\, & false negative rate \\[1pt]  \hline
\end{tabular}
\vspace{1mm}
\caption{\textbf{Overview of variables used in testing model.} An
  overview of the main variables and parameters that will be used in
  developing our testing model. The sets $[0,N]$ and $[0,Q]$ contain
  all integers from $0$ up to $Q$ and $N$, respectively. The set $f,
  f_{\rm b}, \tilde{f}_{\rm b}:[0,1]$ denotes all rational numbers
  between 0 and 1. For ${\rm FNR}, {\rm FPR}$, $[0,1]$ represents all
  real numbers between 0 and 1. We assume that $Q$, $Q^{+}$,
  $\tilde{Q}^{+}$, $f_{\rm b}\!\coloneqq Q^{+}/Q$, and $\tilde{f}_{\rm
    b}\!\coloneqq\tilde{Q}^{+}/Q$ are determined by testing a
  population of known size $N$.}
\label{tab:testing_variables}
\end{table}

Suppose we randomly administer $Q$ tests within a particular short
time period ({\it e.g.}, within one day or one week) to a total
effective population of $N$ previously untested individuals. This
population of individuals is comprised of $S$ susceptible, $I$
infected, and $R$ removed (\ie, recovered or deceased) individuals,
which are unknown. $S$, $I$, and $R$ can dynamically change from one
testing period to another due to transmission and recovery dynamics,
as well as removal from the untested pool by virtue of being
tested. The total population $N=S+I+R$ can also change through
intrinsic population dynamics (birth, death, and immigration), but can
assumed to be constant over the typical time scale of an epidemic that
does not cause mass death.

We start the derivation of our statistical model by first fixing $S$,
$I$ and $R$, assuming both perfect error-free testing, considering a
``testing with replacement'' scenario, in which tested individuals can
be retested within the same time window. Under these conditions, the
probability that $q$ tests are returned positive and $Q^{-} = Q -
q$ tests are returned negative is
\begin{equation}
P_{\rm true}(Q^{+}=q\vert Q, S, I, R) = 
{Q \choose q} {f}^{q}(1-f)^{Q-q},
\label{P_EXACT_BINOMIAL}
\end{equation}
where the parameter
\begin{equation}
f \equiv {(I+R)\over N}\quad \mbox{or} \quad  {I \over N},
\label{F}
\end{equation}
are simply the probabilities of identifying currently and previously infected
individuals with tests such as serological (antibody) tests, or of
detecting current infections with viral load tests, respectively. Note
that testing with replacement renders $P_{\rm true}$ dependent only on
$Q$ and $f$, and not explicitly on $I, S, R$ or $N$. The binomial
expression \ref{P_EXACT_BINOMIAL} is accurate when the number of tests
are much smaller than the population $(I+R)$ or $I$.

Equation \ref{P_EXACT_BINOMIAL} describes perfect error-free and
random testing.  However, if there is some prior suspicion of being
infected, the administration of testing may be biased. For example,
certain jurisdictions focus testing primarily on hospitalized patients
and people with significant symptoms~\cite{fenton2020covid}, thus
biasing the tests to those that are infected. We quantify such testing
biases through a biased-testing function $B(Q^+)\in \mathbb{R}_{\geq
  0}$, leading to the following modification of
Eq.~\ref{P_EXACT_BINOMIAL}:

\begin{equation}
P_{\rm true}(Q^{+}=q\vert Q, f) = 
{{Q \choose q} f^{q}(1-f)^{Q-q} B(q) \over 
\sum_{k=0}^{Q}{Q \choose k} f^{k}(1-f)^{Q-k} B(k)}.
\label{P_EXACT_BINOMIAL_BIAS}
\end{equation}
We define the biased-testing function $B(q)$ as a weighting over
certain numbers $q$ of positive tests.  A convenient choice of this
weighting is a ``Boltzmann'' functional form $B(q) =
e^{qU_{I}}e^{(Q-q)U_{S}}$, for which $P_{\rm true}(Q^{+}=q\vert
f, Q)$ becomes

\begin{align}
P_{\rm true}(Q^{+}=q\vert Q, f, b) & = 
{{Q \choose q}f^{q}(1-f)^{Q-q} e^{qb} \over 
\sum_{k=0}^{Q}{Q \choose k} f^{k}(1-f)^{Q-k} e^{kb}} \nonumber \\
\: & = {{Q \choose q}(f e^{b})^{q}(1-f)^{Q-q}\over 
[1+(e^{b}-1)f]^{Q}},
\label{P_EXACT_BINOMIAL_U}
\end{align}
where $-U_{I},-U_{S}$ represent ``costs'' for testing infecteds,
susceptibles, and $b\equiv U_{I}-U_{S}\in\mathbb{R}$ is a testing-bias
parameter, which we employ below.

A Gaussian approximation to Eq.~\ref{P_EXACT_BINOMIAL_U} can be found
through the mean and variance of $Q^{+}$:
\begin{align}
P_{\rm true}(Q^{+}=q\vert Q, f, b) & \approx 
{1\over \sigma \sqrt{2\pi}}\exp \left[-{(q-\mu(f,b))^{2}\over 2\sigma^{2}(f,b)}\right],
\label{eq:approximation_w_repl_Qp}
\end{align}
where
\begin{align}
\mu(f,b) & \equiv {Q f e^{b} \over 1+f (e^{b}-1)}\quad\text{and}\quad
\sigma^{2}(f,b) \equiv \mu(f,b) \left(1-{\mu(f,b) \over Q}\right).
\label{eq:mu_sigma_w_repl_Qp}
\end{align}
In addition to describing the probability distribution in
  Eq.~\ref{eq:approximation_w_repl_Qp} as a function of the number of
  positive tests $Q^{+}$, we can also express it in terms of the
  observed positive (and potentially sample-biased) testing fraction
  $f_{\rm b}\coloneqq Q^{+}/Q$:

\begin{align}
P_{\rm true}(f_{\rm b}=x \vert Q, f, b) & \approx 
{1\over \bar{\sigma}(f,b) \sqrt{2\pi}}\exp 
\left[-{(x-\bar{\mu}(f,b))^{2}\over 2\bar{\sigma}^{2}(f,b)}\right],
\label{eq:approximation_w_repl}
\end{align}
where here,

\begin{align}
\bar{\mu}(f,b) & \equiv \mu/Q={f e^{b} \over 1+f(e^{b}-1)} 
\quad \text{and}\quad \bar{\sigma}^{2}(f,b) \equiv \bar{\mu}(f,b) 
\left(1-\bar{\mu}(f,b)\right)/Q.
\label{eq:mu_b}
\end{align}
The expected value of $f_{\rm b}$, $\bar{\mu}(f,b)$, can be understood
as a product of the true underlying infected fraction $f$ and
a bias function $\tilde{B}(f,b)$ that depends on $f$ and the bias
parameter $b$, \ie, $\bar{\mu}(f,b)=f \tilde{B}(f,b)$, where
$\tilde{B}$ is given by
\begin{equation}
\tilde{B}(f,b) = \frac{e^b}{1+f (e^b-1)}.
\label{eq:bias_function}
\end{equation}
Note, that Eq.\,\ref{eq:bias_function} implies that when $b>0$, the
currently and/or previously infected population is favored to be
tested, while for $b<0$, the non-infected and/or susceptible
population is favored. The limits of $b \to \pm \infty$ indicate
testing that is completely biased such that only infected and
susceptible individuals are tested, respectively. Realistic
  values of our bias parameter $b$ are positive and $ \sim O(1)$.

\begin{figure}
    \centering
    \includegraphics[width = \textwidth]{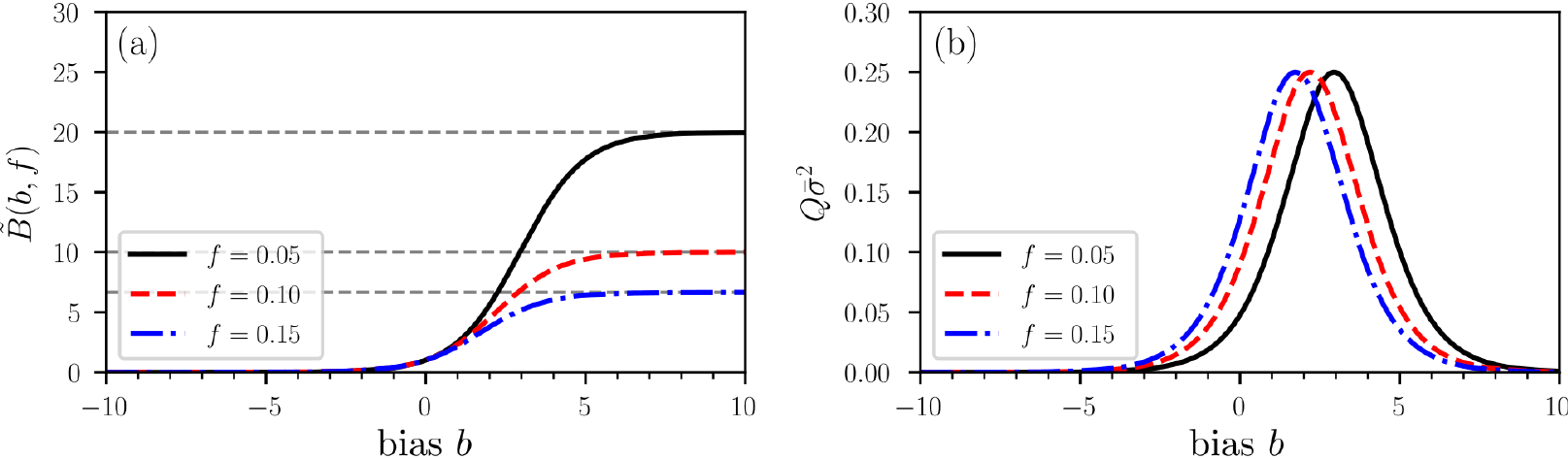}
    \caption{\textbf{Illustration of a bias function
        $\tilde{B}(f,b)$.} (a) The bias function
        $\tilde{B}(f,b)$ (Eq.~\ref{eq:bias_function}) for three
        different fractions $f$ of currently (and previously) infected
        individuals. Grey dashed lines indicate the asymptotic value
        $\tilde{B}(\infty,f)=1/f$. A value $b>0$ indicates a testing
        bias towards currently and/or previously infected individuals
        while susceptible and/or non-infectious individuals are
        preferentially tested for $b<0$. Unbiased testing corresponds
        to $b=0$ and $\tilde{B}(0,f)=1$. (b) The variance
        $\bar{\sigma}^{2}$ exhibits a maximum value of $1/(4Q)$ at a
        typical value of bias $b^{*} = \ln[(1-f)/f]$.}
    \label{fig:bias_function}
\end{figure}

Fig.~\ref{fig:bias_function}(a) shows the bias function
  \ref{eq:bias_function} as a function of $b$ for different infection
  fractions $f$. For $\tilde{B}(f,b) > 1$, the biased-testing fraction
  $f \tilde{B}$ is larger than the unbiased-testing fraction $f$. The
  opposite holds for $\tilde{B}(f,b) < 1$. The variance
  $\bar{\sigma}^{2}$ is plotted as a function of $b$ in
  Fig.~\ref{fig:bias_function}(b) and exhibits a maximum value of
  $1/(4Q)$ at $b^{*} = \ln[(1-f)/f]$.

The probabilities $P_{\rm true}$ derived in
Eqs.\,\ref{P_EXACT_BINOMIAL} and \ref{P_EXACT_BINOMIAL_BIAS}
correspond to ``testing with replacement''. The opposite limit
  is ``testing without replacement''; once an individual is tested
  they are labeled as such and removed from the pool of test targets,
  at least within the specified testing period. This concept of
  sampling with and without replacement commonly arises in the
  measurement of diversity in ecological settings \cite{DIVERSITY}.
  Without replacement, and still under conditions of perfect random
  testing, two slightly different forms for $P_{\rm true}$ arise for
  the different type of tests (\textit{e.g.}, antibody \textit{vs}
  PCR/viral load). For antibody tests that perfectly identifies
  recovered (or deceased) individuals as being previously infected,
  Eq.~\ref{P_EXACT_BINOMIAL} is replaced by
\begin{align}
P_{\rm true}(Q^{+}=q\vert S, I, R, Q) &  = 
{\displaystyle{I + R \choose q}{S \choose Q-q}\over {\displaystyle{N \choose Q}}}
\nonumber \\
\: & = {Q \choose q}\prod_{n=0}^{Q-1}\left({1\over N-n}\right)
\prod_{i=0}^{q-1}\!(I+R-i)\!\!\prod_{j=0}^{Q-q-1}\!\!\!(S-j),
\label{PA_EXACT}
\end{align}
where the binomial coefficients for $q=0,Q$ are zero.  On the other
hand, if the perfect test only identifies individuals that currently
have a viral load, the susceptible and recovered (or deceased)
individuals both test negative and $P_{\rm true}$ is described by
\begin{align}
P_{\rm true}(Q^{+}=q\vert S, I, R, Q) & = \displaystyle
{\displaystyle{I \choose q}{S+R \choose Q-q}\over {\displaystyle{N \choose Q}}}
\nonumber \\
\: & \displaystyle = {Q \choose q}\prod_{n=0}^{Q-1}\left({1\over N-n}\right)
\prod_{i=0}^{q-1}(I-i)\!\!\prod_{j=0}^{Q-q-1}\!\!\!(S+R-j).
\label{PV_EXACT}
\end{align}
The expressions for $P_{\rm true}$ when tested subjects are not
replaced, unlike in the case of testing with replacement, depend
explicitly on $S, I, R$, and $N$.

To incorporate testing bias in into the probabilities $P_{\rm true}$
for testing without replacement, first consider Eqs.\,\ref{PV_EXACT}
and think of ${I \choose q} {S+R \choose Q-q}$ as the number of
ways of distributing $q$ positive tests among $I$ infected
individuals, and $Q-q$ negative tests among $S+R$ uninfected
individuals.  As in the biased-testing formulation of
Eq.~\ref{P_EXACT_BINOMIAL_BIAS}, we interpret the bias as a factor
$B(q)$ that weights more tests in $I$ or $S+R$ pools:
\begin{equation}
P_{\rm true}(Q^{+}=q\vert Q,S,I,R, B) \equiv 
{{I  \choose q}{S+R\choose Q-q} B(q)\over 
\sum_{k=0}^{Q}{I \choose k}{N-I \choose Q-k}B(k)}.
\label{WEIGHT}
\end{equation}
To obtain the ``testing without replacement'' equivalent of
Eq.~\ref{P_EXACT_BINOMIAL_U}, we again set $B(q) =
e^{qU_{I}}e^{(Q-q)U_{S}}=e^{bq} e^{Q U_S}$. By using the
Chu--Vandermonde identity
\begin{equation}
\sum_{k=0}^{Q}{I \choose k}{N-I \choose Q-k} = {N \choose  Q},
\end{equation}
we verify that when $b=0$ the expression \ref{WEIGHT} reduces to
Eq.~\ref{PV_EXACT}.

For the exponential form of $B$, the normalizing ``partition
function'' becomes
\begin{equation}
\sum_{k=0}^{Q}{I \choose k}{N-I \choose Q-k}e^{kb}
= {N-I \choose Q}{_2}F_{1}(-I, -Q; N-I-Q+1; e^{b}),
\end{equation}
where ${}_2F_1$ denotes the (ordinary) hypergeometric function. 
Thus, the distribution of positive tests $Q^+$ under biased testing without
replacement for viral load-type tests can thus be expressed as
\begin{equation}
P_{\rm true}(Q^{+}=q\vert Q,S,I,R,b) \equiv 
{e^{qb}  {Q \choose q}
\prod_{k=0}^{q-1}\left({I - k \over S+R-Q^{-}-k}\right)   
\over {_2}F_{1}(-I, -Q; N-I-Q+1; e^{b})}
%
%
\label{WEIGHT2}
\end{equation}
The distribution of positives under biased testing without replacement
for antibody-type tests becomes (analogous to Eq.\,\ref{PA_EXACT}) is
\begin{equation}
P_{\rm true}(Q^{+}=q\vert Q, S,I,R, b) \equiv 
{e^{qb}{Q \choose q}
\prod_{k=0}^{q-1}\left({I+R - k \over S-Q^{-}-k}\right)
\over {_2}F_{1}(-I-R, -Q; N-I-R-Q+1; e^{b})}
%
%
\label{WEIGHT3}
\end{equation}
where $Q^{-}\equiv Q-q$. The above two expressions are
  equivalent except for the merging of the $R$ pool with uninfected
  susceptibles $S$ in one case, or with current infecteds $I$ in the
  other. Because the testing decreases with the number of tests
  administered, Eqs.~\ref{WEIGHT2} and \ref{WEIGHT3} cannot be reduced
  to functions of a simple positive-test fraction $f_{\rm b}$ or to
  universally accurate simple Gaussian forms.

\subsection{Testing errors}

The probability distributions $P_{\rm true}$ that we derived in
Eq.\,\ref{P_EXACT_BINOMIAL_BIAS} and in
Eqs.\,\ref{PA_EXACT}--\ref{PV_EXACT} assume that testing is
error-free, \ie, that the false-negative rate $ {\rm FNR} = 1 - {\rm
  TPR}=0$ and false-positive rate ${\rm FPR} = 1- {\rm TNR} =0$, or
equivalently that the true-positive rate ${\rm TPR} = 1$ and the
true-negative rate ${\rm TNR}=1$. To incorporate erroneous
  testing, we now construct the probability distribution of
  error-generated deviation $P_{\rm err}(\tilde{Q}^{+}=q\vert
  Q^{+}=q')$ over the number of ``apparent'' positives $q$ from tests that
  carry nonzero FPRs and FNRs, given that $q'$ positives would be
  recorded if the tests were perfect. If $q$ apparent positive tests
are tallied, $p$ of them might have been true positives drawn from the
perfect-test positives $q'$ in ${q'\choose p}$ ways, while the
remaining $k=q-p$ apparent positives might have been erroneously
counted as positives drawn from the $Q^{-}\equiv Q-q'$ true
negatives. The remaining $q'-p$ true positive tests might have been
erroneously tallied as false negatives, while the remaining $Q^{-}-k$
negative tests might have been correctly tallied as true negatives.
Assuming nonzero ${\rm FPR}$ and ${\rm FNR}$, we find that the
probability distribution of finding $0\leq q \leq Q$ apparent positive
tests is
\begin{align}
& P_{\rm err}(\tilde{Q}^{+}=q\vert Q^{+}=q', Q, {\rm FPR},{\rm FNR}) = \nonumber \\
& \hspace{1.5cm} \sum_{p=0}^{q}
{q' \choose p} ({\rm TPR})^{p} ({\rm FNR})^{q'-p}
{Q-q' \choose q-p} ({\rm FPR})^{q-p}({\rm TNR})^{Q-q'-(q-p)},
\label{PM1}
\end{align}
where we invoked the identities TPR + FNR = 1 and FPR + TNR = 1.  The
total error-prone distribution $P_{\rm TOT}(\tilde{Q}^{+}=q\vert
Q,S,I, R,b)$ of recording $q$ positives after having administered $Q$
tests under bias and/or testing errors is given by convolving the
probability $P_{\rm err}$ of finding $q$ apparent tests given $q'$
true positive tests with the probability $P_{\rm true}$ of finding
$q'$ positive tests under perfect testing
(Eq.\,\ref{P_EXACT_BINOMIAL_U} or Eq.\,\ref{WEIGHT2}):
\begin{align}
&P_{\rm TOT}(\tilde{Q}^{+}=q\vert Q,S,I, R, b, {\rm FPR}, {\rm FNR})
= \nonumber \\
& \hspace{1.5cm} \sum_{q'=0}^{Q} P_{\rm err}(\tilde{Q}^{+}=q\vert  Q^{+}=q',Q, 
{\rm FPR},{\rm FNR}) P_{\rm true}(Q^{+}=q'\vert Q, S,I, R,b).
\label{FULL_P}
\end{align}

This convolution can be further simplified by taking the Gaussian
limit of the binomial distributions that appear in $P_{\rm err}$ and
in $P_{\rm true}$. To be concrete, we use $P_{\rm true}$ as written in
Eq.\,\ref{P_EXACT_BINOMIAL_BIAS} under the testing with replacement
scenario and use the approximation
\ref{eq:approximation_w_repl_Qp}. The same Gaussian
  approximation can be used for all terms in $P_{\rm err}$ from
  Eq.\,\ref{PM1}. After approximating the summation over $p\in (0,q)$
  in Eq.~\ref{PM1} by an integral over $p\in (-\infty,\infty)$
  (provided $q$ is sufficiently large), we find
\begin{align}
P_{\rm err}(\tilde{Q}^{+}=q\vert Q, Q^{+}=q', {\rm FPR}, {\rm FNR}) & 
%
%
 \approx {\exp\left[-\displaystyle{{(q-q'{\rm TPR} -(Q-q'){\rm
           FPR})^{2}\over 2q'{\rm FNR}\, {\rm TPR} +2(Q-q'){\rm FPR}\,
         {\rm TNR}}}\right]\over \sqrt{2\pi}\sqrt{q'{\rm FNR}\, {\rm
       TPR} +(Q-q'){\rm FPR}\, {\rm TNR}}}.
\label{Perrapprox}
\end{align}
We now convolve Eq.~\ref{eq:approximation_w_repl_Qp} with
Eq.\,\ref{Perrapprox} as prescribed by Eq.\,\ref{FULL_P} to find
\begin{equation}
P_{\rm TOT}(\tilde{Q}^{+}=q\vert Q, f, b, {\rm FPR}, {\rm FNR}) \approx
{1 \over \sigma_{\rm T}\sqrt{2\pi}} 
\exp\left[-{(q-\mu_{\rm T})^{2} \over 2\sigma_{\rm T}^{2}} \right],
\end{equation}
where
\begin{align}
\begin{split}
\mu_{\rm T}(f,b,{\rm FPR}, {\rm FNR}) \equiv & 
Q\left[\bar{\mu}(f,b)(1-{\rm FNR})+(1-\bar{\mu}(f,b)){\rm FPR}\right], \\
\sigma_{\rm T}^{2}(f,b,{\rm FPR}, {\rm FNR})
\equiv & 
Q(1-\bar{\mu}(f,b)){\rm FPR}(1-{\rm FPR}) + Q\bar{\mu}(f,b){\rm FNR}(1-{\rm FNR}) \\
\: &  \hspace{2cm} +Q\bar{\mu}(f,b)(1-\bar{\mu}(f,b))(1-{\rm FNR}-{\rm FPR})^{2}.
\end{split}
\end{align}
with $\bar{\mu}(f,b)$ the mean value of the fraction of observed positives 
under biased, but perfect testing (see Eq.~\ref{eq:mu_b}).
\begin{figure}
\includegraphics[width = \textwidth]{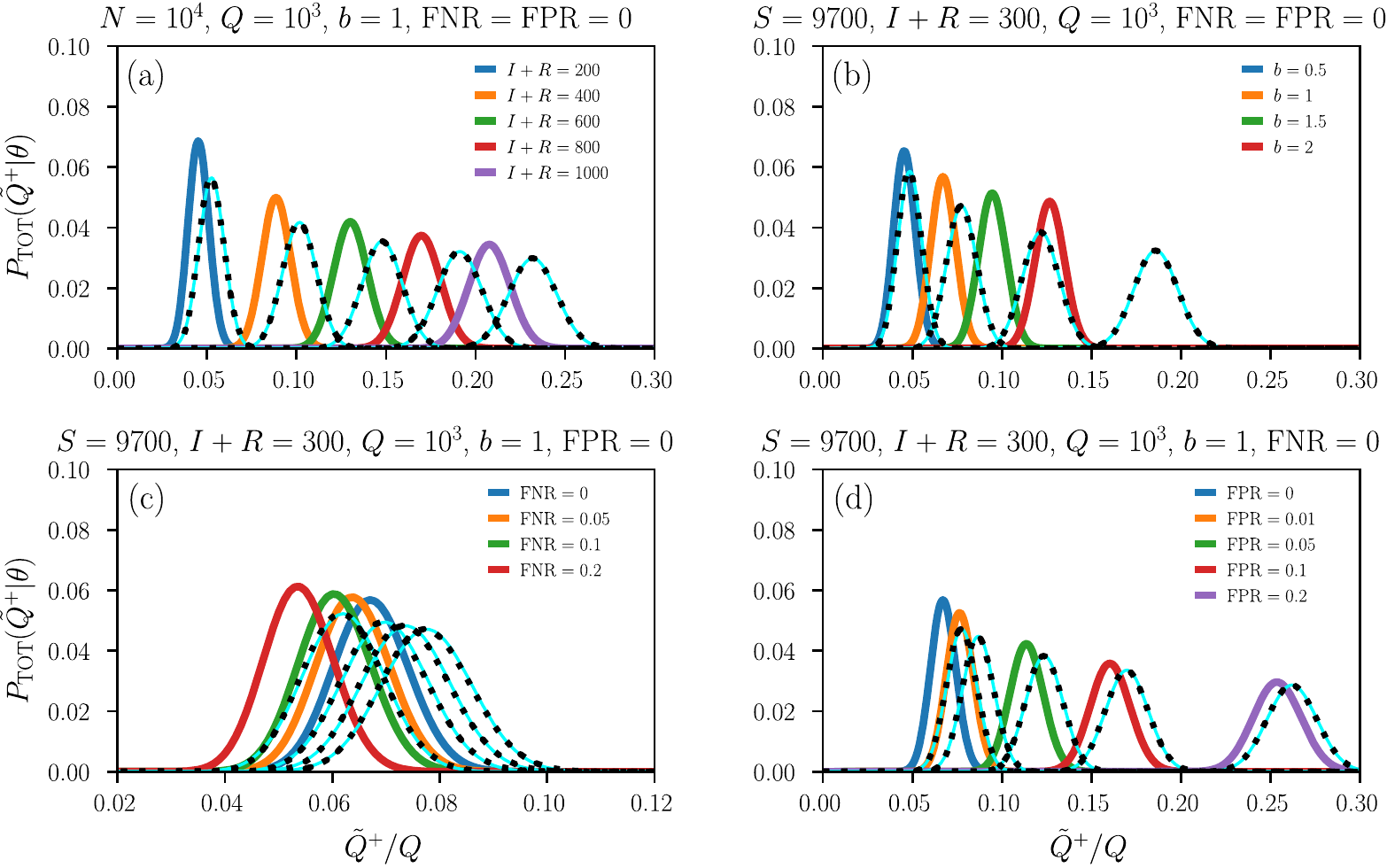}
\caption{\textbf{Distribution of apparently positive tests.} Plots of
  $P_{\rm TOT}(\tilde{Q}^{+}=q\vert Q,I, S, R,b)$ with $N=S+I+R=10^4$,
  $Q=10^3$, and different (a) values of $I+R$, (b) testing biases $b$,
  (c) FNRs, and (d) FPRs. The Gaussian approximation (solid light blue
  lines) of Eq.~\ref{eq:P_TOT_APPROX} provides an accurate
  approximation of $P_{\rm TOT}$. Dashed black lines correspond to
  distributions with replacement and the remaining solid colored lines
  correspond to those without replacement.}
\label{fig:testing}
\end{figure}
The mean number of apparent positive tests $\mu_{\rm T}$ is given
by the sum of the expected value of true positive tests (\ie, $Q
\bar{\mu}(f,b) (1-\mathrm{FNR})=Q \bar{\mu}(f,b)\mathrm{TPR}$) and the
expected value of false positive tests (\ie, $Q(1-\bar{\mu}(f,b))
\mathrm{FPR}$).  Based on the derived expressions for $\mu_{\rm T}$
and $\sigma_{\rm T}^2$, we define the random variable as the fraction
of observed positive tests $\tilde{f}_{\rm b} \equiv \tilde{Q}^{+}/Q$
under biased and error-prone testing and obtain
\begin{equation}
P_{\rm TOT}(\tilde{f}_{\rm b}=x\vert Q, f, b, {\rm FPR}, {\rm FNR})
\approx {e^{-(x-\bar{\mu}_{\rm T})^{2}/(2\bar{\sigma}_{\rm
      T}^{2})}\over \bar{\sigma}_{\rm T} \sqrt{2\pi}},
\label{eq:P_TOT_APPROX}
\end{equation}
where 
\begin{align}
\begin{split}
\bar{\mu}_{\rm T}(f,b,{\rm FPR}, {\rm FNR}) \equiv & \left[\bar{\mu}(f,b)
(1-{\rm FNR})+(1-\bar{\mu}(f,b)){\rm FPR}\right], \\
Q\bar{\sigma}_{\rm T}^{2}(f,b,{\rm FPR}, {\rm FNR}) 
\equiv & (1-\bar{\mu}(f,b)){\rm FPR}(1-{\rm FPR}) + 
\bar{\mu}(f,b){\rm FNR}(1-{\rm FNR}) \\
\: & \hspace{1.8cm} + \bar{\mu}(f,b)(1-\bar{\mu}(f,b))(1-{\rm FNR}-{\rm FPR})^{2}.
\label{eq:mu_sigma_approx}
\end{split}
\end{align}
The Gaussian approximation \ref{eq:mu_sigma_approx} is quite accurate
provided that (i) the number of positive and apparent positive
tests, $Q^+ = q'$ and $\tilde{Q}^+ = q$, are sufficiently large, and (ii), the
quantities $\bar{\mu}$ , ${\rm FNR}$, and ${\rm FPR}$ are not too
close to 0 or 1. In Table~\ref{tab:testing_variables}, we provide an
overview of the main variables used in the statistical testing model
\eqref{PM1}--\eqref{eq:mu_sigma_approx}.

Fig.~\ref{fig:testing} shows the distribution of apparently infected
individuals $P_{\rm TOT}$ for different numbers of infected/recovered
individuals [Fig.~\ref{fig:testing} (a)], testing biases
[Fig.~\ref{fig:testing} (b)], and testing sensitivities (\ie, true
positive rates, ${\rm TPR}$s) and specificities (\ie, true negative
rates, ${\rm TNR}$s) [Fig.~\ref{fig:testing} (c--d)]. Solid light blue
lines represent the Gaussian approximation \ref{eq:P_TOT_APPROX} and
dashed black lines and the remaining colored lines are calculated by
directly evaluating Eq.~\ref{FULL_P} with replacement
(Eq.~\ref{P_EXACT_BINOMIAL_BIAS}) and without replacement
(Eq.~\ref{WEIGHT3}), respectively. The ${\rm FNR}$s that we consider
in Fig.~\ref{fig:testing} (c) are chosen in accordance with reported
sensitivities of serological and RT-PCR tests for
SARS-CoV-2~\cite{lassauniere2020evaluation,whitman2020test,fang2020sensitivity,wang2020detection}. We
observe that an increase in the ${\rm FNR}$ slightly shifts the
distribution $P_{\rm TOT}$ towards smaller values of apparently
infected individuals, which is consistent with the $\mathrm{FNR}$
dependence of the mean $\bar{\mu}_{\rm T}$
(Eq.~\ref{eq:mu_sigma_approx}). For serological and RT-PCR tests, the
$\mathrm{FPR}=1-\mathrm{TNR}$ is about 5\%. A smaller specificity
would lead to larger ${\rm FPR}$s and a shift of $P_{\rm TOT}$ towards
larger values of $\tilde{Q}^{+}$ and $\tilde{f}_{\rm b}$
[Fig.~\ref{fig:testing} (d)]. Our results show that $P_{\rm TOT}$ is
more affected by variations in the testing specificity than by
variations in testing sensitivity.

\subsection{Temporal variations and test heterogeneity}
Up to now, we have discussed single viral-load and antibody tests
(with and without replacement) but have not considered temporal
variations in the number of tests $Q$, the number of returned
  positives $\tilde{Q}^{+}$, and heterogeneity in $\mathrm{FNR}$ and
$\mathrm{FPR}$ that are associated with different classes (types,
manufacturing batches, etc.) of assays.  To make our model applicable
to empirical time-varying testing data, we use $S_t$, $I_t$, $R_t$ to
denote the number of susceptible, infected, and removed individuals at
time $t$ (or in successive time windows labeled by $t$),
respectively. If $K>1$ test classes are present, we also include an
additional index $c\in\{1,\dots,K\}$ in all relevant model
parameters. The testing bias and the total number of tests may be both
test-class and time-dependent. That is, $b=b_{t,c}$ and
$Q=Q_{t,c}$. Test specificity and sensitivity mainly depend on the
assay type and not on time. We thus set
$\mathrm{FPR}=\mathrm{FPR}_{c}$ and $\mathrm{FNR}=\mathrm{FNR}_{c}$.
\section{Inference of prevalence and application to COVID-19 data}
\label{sec:covid_data}
\begin{figure}
    \centering
    \includegraphics[width=\textwidth]{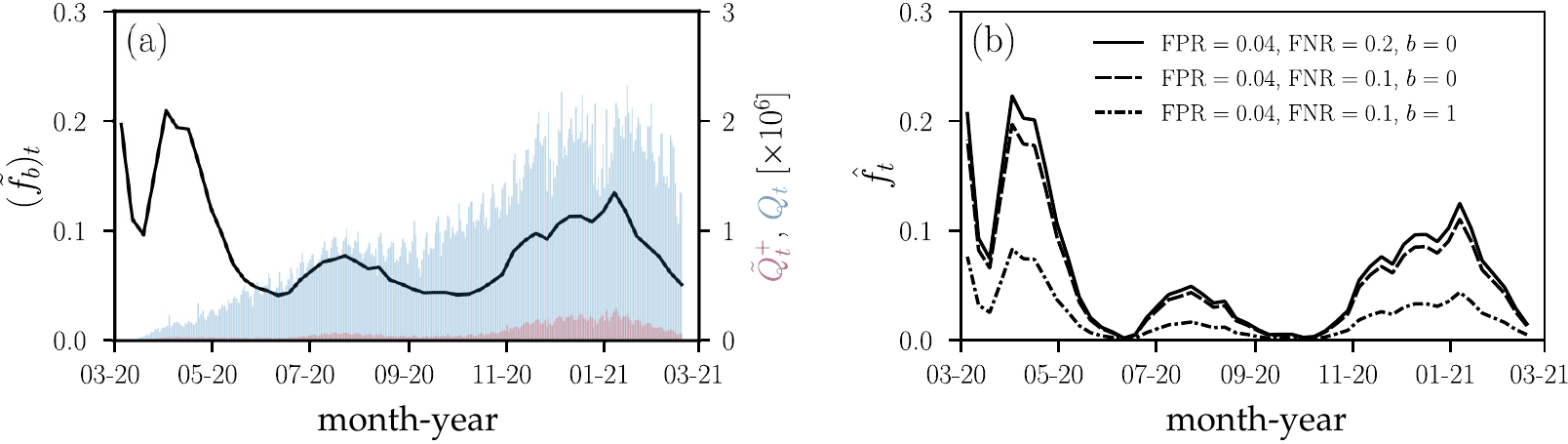}
    \caption{\textbf{Observed and corrected proportions of positive
        tests in the United States.} (a) The solid black line
      represents the 7-day average of the proportion of positive tests
      $(\tilde{f}_{\rm b})_t=\tilde{Q}_t^+/Q_t$ in the United
      States. Blue and red bars show the corresponding total number of
      daily tests $Q$ and apparent positive tests $\tilde{Q}^{+}_t$,
      respectively. (b) The corrected proportion of positive tests
      $\hat{f}_t$, found by inverting Eq.~\ref{eq:ML_estimate}, for
      different $\mathrm{FPR}$, $\mathrm{FNR}$, and bias
      combinations.}
    \label{fig:testing_rate}
\end{figure}
For a certain time window, one often wishes to infer $I_t+R_t$ and
$S_t$, or $I_t$ and $S_t+R_t$ from values of $b_{t,c}$, $Q_{t,c}$, and
$q_{t,c}$. Alternatively, since $b_{t,c}$ is difficult to
independently ascertain, one may only be able to infer $(f_b)_{t,c} =
f(S_t, I_t, R_t, b_{t,c})$. For a single test result $\tilde{q}_{t,c}$
(or $(\tilde{f}_{\rm b})_{t,c}$), we can generate the maximum
likelihood estimate (MLE) of the bias-modified prevalence
$(\hat{f}_b)_{t,c}$ by setting the measured value $(\tilde{f}_{\rm
  b})_{t,c} = \bar{\mu}_{\rm T}[(\hat{f}_b)_{t,c}]$ to find
\begin{equation}
(\tilde{f}_{\rm b})_{t,c} = (\hat{f}_b)_{t,c}(1-{\rm FNR}_c)+(1-(\hat{f}_b)_{t,c}){\rm FPR}_c
\label{eq:ML_estimate}
\end{equation}
and
\begin{equation}
(\hat{\bar{\sigma}}_{\rm T})_{t,c}^{2} \approx (\bar{\sigma}_{\rm T})_{t,c}^{2}[(\hat{f}_b)_{t,c}].
\end{equation}
Since $(\hat{f}_b)_{t,c}=\hat{f}_{t,c} \tilde{B}(\hat{f}_{t,c},b_{t,c})$,
Eq.~\ref{eq:ML_estimate} can be solved for $\hat{f}_{t,c}$:
\begin{equation}
\hat{f}_{t,c}={(\tilde{f}_{\rm b})_{t,c} 
-{\rm FPR}_c \over e^b[1-{\rm FNR}_c
-(\tilde{f}_{\rm b})_{t,c}]+(\tilde{f}_{\rm b})_{t,c}-{\rm FPR}_c}.
\end{equation}

The posterior distribution $P_{\rm post}$ over values
of $f$ can be found through Bayes' theorem:
\begin{align}
P_{\rm post}(f \vert \tilde{Q}^{+}, Q, {\rm FPR}, {\rm FNR}) = 
{P_{\rm TOT}(\tilde{Q}^{+}\vert Q, f, {\rm FPR}, {\rm FNR})
P_{0}(f) \over 
\int_{0}^{1} P_{\rm TOT}(\tilde{Q}^{+}\vert Q, f',{\rm FPR}, {\rm FNR})
P_{0}(f')\, {\rm d}f'},
\label{eq:Bayes}
\end{align}
where $P_{0}(f)$ is a prior distribution over the underlying infection
fraction $f$ in the population. For notational brevity, we did not
include the indices $c$ and $t$ in Eq.~\ref{eq:Bayes}. We can again
simplify the analysis by using the Gaussian approximation and a simple
initial uniform prior, $P_{0}(f<f_{\rm max}\leq 1) = 1/f_{\max}$.

As an example, we collected US testing data~\cite{CovidDataTracking}
from March 2020 to March 2021. Figure \ref{fig:testing_rate}(a) shows
the daily number of observed positive tests $\tilde{Q}_t^+$ (red bars)
and the corresponding total daily number of tests $Q_t$ (blue
bars). The 7-day average of the observed positive testing rate
$(\tilde{f}_{\rm b})_t=\tilde{Q}_t^+/Q_t$ is indicated by the black solid
line.  The first drop in $(\tilde{f}_{\rm b})_t$ in March 2020 was
associated with the initially very limited number of available
SARS-CoV-2 testing infrastructure followed by the ramping up of testing
capacity. After new cases surged by the end of March and in April
2020, different types of stay-at-home orders and distancing policies
with different durations were implemented across the United
States~\cite{moreland2020timing}. In June and July 2020, reopening
plans were halted and reversed by various jurisdictions to limit the
resurgence of COVID-19~\cite{timeline_covid_us}.

In Fig.~\ref{fig:testing_rate}(b), we show the corrected proportion of
positive tests $\hat{f}_t$, found by numerically inverting
Eq.~\ref{eq:ML_estimate} for different $\mathrm{FPR}$, $\mathrm{FNR}$,
and bias combinations. We observe that a small $\mathrm{FPR}=0.04$
shifts values $(\tilde{f}_{\rm b})_t\approx 0.05$ towards zero such that the
corrected positive testing rate $\hat{f}_t\approx 0$. Reducing the
$\mathrm{FNR}$ from $0.2$ to $0.1$ has only little effect on the
corrected proportion of positive tests $\hat{f}_t$ [solid black and
  dashed lines in Fig.~\ref{fig:testing_rate}(b)]. Accounting for a
positive testing bias of $b=1$ (\ie, preferential testing of infected
and symptomatic individuals by a factor of $e$), however, markedly
changes the inferred $\hat{f}_t$ [dashed-dotted black line in
  Fig.~\ref{fig:testing_rate}(b)].
\section{Inference of bias $b$}
\label{sec:bias}
One way to estimate the testing bias $b$ is to identify a smaller
subset of control tests within a jurisdiction that is believed to be
unbiased and compare it with the reported fraction of positive tests
obtained via standard (potentially biased) testing procedures.

We can derive a rather complete methodology to estimate
  bias by formally comparing the statistics of two sets of tests
  applied to the same population. The first set of control tests with
  testing parameters $\theta_{0} = \{Q_{0}, f, {\rm FPR}_{0}, {\rm
    FNR}_{0}\}$ is known to be unbiased (has prior distribution
  $\delta(b)$), while the second set is taken with known parameters
  $\theta = \{Q, {\rm FPR}, {\rm FNR}\}$, but unknown testing bias
  $b$. For example, the control set may consist of a smaller number
  $Q_{0}$ of tests that are administered completely randomly, while
  the second set may be the scaled-up set of tests with $Q >
  Q_{0}$. Since both sets of tests are applied roughly at the same
  time to the same overall population, the underlying positive
  fraction $f$ is assumed to be the same in both test sets.  We can
  then use Bayes' rule on the first unbiased test set to infer $f$:

\begin{equation}
P_{\rm post}(f \vert \tilde{f}_{0}, \theta_{0}, b=0)=
{P_{\rm TOT}(\tilde{f}_{0}\vert f, \theta_{0}, b=0)P_{0}(f)\over 
\int_{0}^{1} P_{\rm TOT}(\tilde{f}_{0} \vert f', \theta_{0}, b=0)P_{0}(f')\,{\rm d} f'}.
\end{equation}
The probability distribution over $b$ for a specified value of $f$ can also be 
constructed from Bayes' rule

\begin{equation}
P_{\rm post}(b \vert \tilde{f}_{\rm b}, f, \theta)=
{P_{\rm TOT}(\tilde{f}_{\rm b}\vert f, \theta, b)P_{0}(b)\over 
\int_{-\infty}^{\infty} P_{\rm TOT}(\tilde{f}_{\rm b} 
\vert f, \theta, b')P_{0}(b')\,{\rm d}b'},
\end{equation}
where $P_{0}(\cdot)$ are prior distributions over the relevant
parameters. The final distribution over the bias factor, given the two
measurements $\tilde{f}_{0}$ and $\tilde{f}_{\rm b}$ derived from the
two sets of tests with testing parameters $\theta_{0}$ and $\theta$,
can be found using

\begin{equation}
P_{\rm b}(b\vert \tilde{f}_{\rm b}, \tilde{f}_{0}, \theta, \theta_{0}) =
\int_{0}^{1} P_{\rm post}(b \vert \tilde{f}_{\rm b}, f, \theta)P_{\rm
  post}(f \vert \tilde{f}_{0}, \theta_{0}, b=0){\rm d} f.
\end{equation}

Of course, a simpler maximum likelihood estimate can also be
  applied to data by first inferring the most likely value of $f$ from
  the control test set. We can use the number of positive tests in the
  control sample $Q^{+}_{0}$ to define the variable $\tilde{f}_{0} =
  Q^{+}_{0}/Q_{0}$. One can then maximize $P_{\rm TOT}(\tilde{f}_{0}
  \vert f, \theta_{0}, b=0)$ with respect to $f$ and use this value $\hat{f}$ in $P_{\rm TOT}(\tilde{f}_{\rm b}\vert
\hat{f}, \theta, b)$.  Maximizing $P_{\rm TOT}(\tilde{f}_{\rm b}\vert
\hat{f}, \theta, b)$ with respect to $b$ then gives the MLE estimate
$\hat{b}$. We can use random and unbiased sampling
  results obtained in the German jurisdiction of Gangelt, North
  Rhine-Westphalia~\cite{streeck2020infection}. A total of 600 adult
  persons with different last names were randomly selected from a
  population of 12,597 and asked to participate in the study together
  with their household members. The resulting study comprised of
  $Q_{0}=919$ subjects who underwent serological and PCR testing
  between March 31-April 6, 2020. The specificity and sensitivity
  corrected, unbiased positive test fraction was determined to be $f
  =15.53\%$ (95\% CI; 12.31\%--18.96\%). Thus, we use this value as an
  estimate for the true underlying positivity rate $\hat{f}$. The larger sample taken across North Rhine-Westphalia between March
30--April 5, 2020 was measured ($Q\approx 25,000$) to be
$\tilde{f}_{\rm b} \approx 0.1$~\cite{NRW_Data}. Assuming that this value is
also error-corrected, an estimate of the bias $\hat{b}$ in this main
testing set can be found by solving $\tilde{f}_{\rm b} \approx 0.1
=\bar{\mu}_{\rm T}(\hat{f}=0.1553,\hat{b}, {\rm FPR}={\rm FNR}=0)=
\bar{\mu}(\hat{f}=0.1553, \hat{b})$ for $\hat{b}$.  We find that the
difference between the unbiased positive testing rate of 15.53\% and
10\% corresponds to a bias of $\hat{b}=-0.50$. This negative bias
likely arises because Gangelt was a an infection hotspot within the
entire North Rhine-Westphalia region, so the control sample was
probably not unbiased.  For comparison, a higher biased positive
testing rate of 20\% would lead to an estimated testing bias
$\hat{b}=0.31$.
\begin{figure}
    \centering
    \includegraphics[width = 0.55\textwidth]{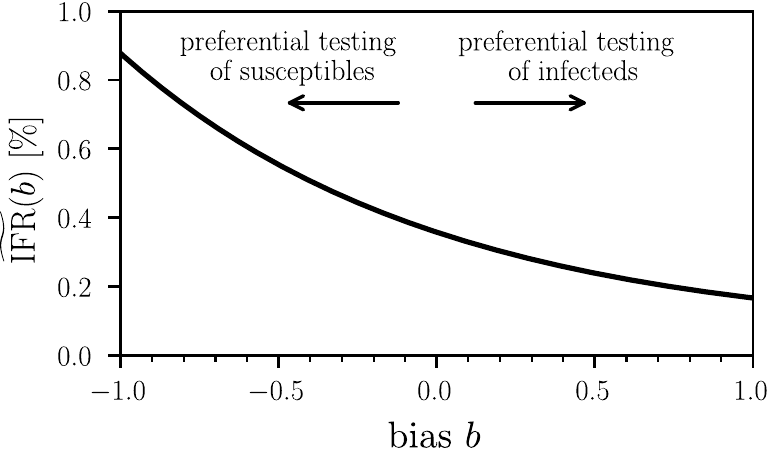}
    \caption{\textbf{Dependence of the observed IFR on testing bias.}
      The observed infection fatality ratio $\widetilde{\rm IFR}$
      (Eq.~\eqref{eq:observed_IFR}) as a function of testing bias
      $b$. We used the example of the German jurisdiction Gangelt and
      set $\hat{f}=0.1553$, $D=7$, and $N=\text{12,597}$. A value $b>0$
      indicates a testing bias towards currently and/or previously
      infected individuals while susceptible and/or non-infectious
      individuals are preferentially tested for $b<0$. Unbiased
      testing corresponds to $b=0$.}
    \label{fig:IFR_bias}
\end{figure}

The number of total deaths on April 6, 2020 amounted to 7. Hence, the
corresponding estimate of the infection fatality ratio (IFR), the
number of disease-induced deaths $D_t$ divided by the total number of
cases $N_t$ at time $t$, in this jurisdiction on April 6, 2020 was
$7/(0.1553\times\text{12,597})=0.36$\% (95\% CI;
0.29\%--0.45\%)~\cite{streeck2020infection}. If only a biased estimate
of the proportion of positive cases is known and not the true value
$f$, we can use our framework to distinguish between the true ${\rm
  IFR}_t=D_t/(N_t-S_t)=D_t/(f_0 N_t)$ and the observed infection
fatality ratio
\begin{equation}
\widetilde{\rm IFR}_t = \frac{D_t}{f N_t}=\frac{D_t}{f \tilde{B}(f,b) N_t}\,.
\label{eq:observed_IFR}
\end{equation}
Figure \ref{fig:IFR_bias} shows the observed $\widetilde{\rm IFR}$ as
a function of testing bias $b$ for the aforementioned example of the
German jurisdiction of Gangelt. Values of $b>0$ correspond to
preferential testing of infected individuals and thus lead to an
apparently lower $\widetilde{\rm IFR}$. The opposite holds for $b<0$.

\section{Summary and Conclusions}
\label{sec:conclusion}
Radiological testing methods such as chest computed tomography (CT)
are used sporadically to identify COVID-19-induced pneumonia in
patients with negative tests~\cite{xie2020chest}, However, the
overwhelming majority of COVID-19 tests are based on serological (or
antibody) tests and rapid antigen tests, ELISA, and RT-PCR
assay~\cite{CDC_tests_overview}. These tests are designed to
subsequently output a binary signal, either infected or not. The
population statistics of this output are affected by testing errors
and bias. False positive and false negative rates of serological tests
are generally smaller than those of rapid antigen tests and RT-PCR
tests. However, serological tests are unable to identify early-stage
infections since they are measuring antibody titres that usually
develop a few days up to a few weeks after infection. In addition to
the occurrence of false positives and false negatives (\ie, type-I and
type-II errors), certain demographic groups (\eg, elderly people or
those with comorbidities such as heart and lung diseases) may be
overrepresented in testing statistics.

To quantify the impact of both type-I/II errors and testing bias on
reported COVID-19 case and death numbers, we developed a mathematical
framework that describes erroneous and biased sampling (both with and
without replacement) from a population of susceptible, infected, and
removed (\ie, recovered or deceased) individuals. We identify a
positive testing bias $b>0$ with an overrepresentation of previously
or currently infected individuals in the study population. Conversely,
a negative testing bias $b<0$ corresponds to an overrepresentation of
susceptible and/or non-infectious individuals in the study
population. We derived maximum likelihood estimates of the
testing-error and testing-bias-corrected fraction of positive
tests. Our methods can be also applied to infer the full distribution
of corrected positive testing rates over time and for different types
of tests across different jurisdictions.

The mathematical quantity that underlies most of our analysis is the
proportion of apparent positive tests.  As pointed out
in~\cite{omori2020changes}, the \emph{absolute} number of positive
tests may not capture the actual growth of an epidemic due to
limitations in testing capacity. Still, many jurisdictions report
absolute case numbers without specifying the total number of tests or
additional information about test type, date of test, and duplicate
tests~\cite{wsj_positivity_rate}, rendering interpretation and
application to epidemic surveillance challenging. For a reliable
picture of COVID-19 case numbers, more complete testing data,
including total number of tests, number of positive tests, test type,
and date of test, has to be reported and made publicly available at
online data repositories. To correct for false positive, false
negatives, and testing bias in testing statistics
(Fig.~\ref{fig:testing_rate}), it will be also important to further
improve estimates of ${\rm FPR}$, ${\rm FNR}$, and $b$ through in
field studies. In particular, estimating the testing bias $b$ requires
random sampling studies similar to that carried out in
\cite{streeck2020infection}. Finally, while we have presented our
analysis in the context of the COVID-19 pandemic, the general results
presented in this paper apply to testing and estimation of severity of
any infectious disease afflicting a population.

\vskip6pt

\enlargethispage{20pt}

\dataccess{Our source codes are publicly available at \url{https://github.com/lubo93/disease-testing}.}

\aucontribute{LB, MRD, and TC contributed equally to the study design, data analyses, and manuscript writing.}

\competing{The authors declare no competing interests.}

\funding{LB acknowledges financial support from the Swiss National
  Fund (P2EZP2\_191888). The authors also acknowledge financial support
  from the Army Research Office (W911NF-18-1-0345), the NIH
  (R01HL146552), and the National Science Foundation (DMS-1814364,
  DMS-1814090).}



\bibliographystyle{IEEEtran}
\bibliography{refs.bib}
\end{document}